\documentclass[twocolumn]{aastex631}

\received{January 19, 2021}
\revised{May 29, 2021}
\accepted{May 31, 2021}

\submitjournal{ApJ}

\shorttitle{Magnetic Fields in Black Hole Binaries}
\shortauthors{Wallace \& Pe'er}

\begin{document}

\title{An Observational Signature of Sub-Equipartition Magnetic Fields in the
Spectra of Black Hole Binaries}

\correspondingauthor{John Wallace}
\email{john.wallace@biu.ac.il, asaf.peer@biu.ac.il}

\author[0000-0001-6180-801X]{John Wallace}
\affiliation{Department of Physics, Bar-Ilan University, Ramat-Gan 52900,
Israel}

\author[0000-0001-8667-0889]{Asaf Pe'er}
\affiliation{Department of Physics, Bar-Ilan University, Ramat-Gan 52900,
Israel}

\begin{abstract}

A common assumption used in the study of accretion disks is that the magnetic
energy density and the kinetic energy density should be in equipartition. This
assumption relies on the faster growth rate of the magnetic field strength
against the kinetic energy of the particles in the flow, for decreasing radius,
combined with a dissipation mechanism that tends towards equipartition. In this
paper, we examine this assumption by modeling the radio, mm and optical spectra
of several black hole binaries in their quiescent state. We use a standard
two-component disk model, consisting of an inner geometrically thick and
optically thin disk, emitting thermal synchrotron radiation, along with an
outer, thin disk, which radiates as a multicolor blackbody. We find that at the
low accretion rates typical of the quiescent state, the spectral shape is
qualitatively reproduced using magnetic fields that are between 0.1\% and 1\% of
the equipartition value, considerably smaller than previously thought. We
discuss our findings in view of (1) the launching of jets in these objects,
which is commonly believed to rely on the presence of a strong magnetic field in
the central region of the disk; and (2) the role of magnetic dissipation in the
structure of the inflow.

\end{abstract}


\section{Introduction} \label{sec:intro}

Black hole binaries (BHBs) exhibit a number of distinct spectral states, which
are classically defined by the hardness of the X-ray emission and the bolometric
luminosity of the system in each state. The different states are in turn
associated with separate modes of accretion. These are generally divided into
four main states as follows, from higher accretion rates to lower: The high/soft
state (HSS), very high state/intermediate state (VHS/IS), low/hard state (LHS)
and quiescent state (QS)
\citep{Esin1997,Zdziarski2004,Remillard2006,Narayan2008}.

The HSS, corresponding to accretion rates $\dot{M} \gtrsim
0.1\dot{M}_\mathrm{Edd}$, is a high luminosity state, consisting of a cool,
geometrically thin disk extending down to the innermost stable circular orbit
(ISCO) \citep{Shakura1973,Esin1997,Remillard2006,Narayan2008}. Such a disk is
optically thick and radiates as a multicolor blackbody disk (MCD) from its
surface \citep{Frank2002}.

At lower accretion rates, the inner regions of the disk do not radiate
efficiently, resulting in a flow which is advection dominated, hot and optically
very thin \citep{Narayan1994}. The inner regions of the disk begin to
``inflate'', resulting in a roughly spherical flow, extending from the ISCO out
to some radius $r_\mathrm{tr}$, the truncation radius, where the flow can once
again radiate efficiently and return to the thin disk structure
\citep{Narayan1996,Esin1997,Homan2005}. For accretion rates $\dot{M} \lesssim
0.1\dot{M}_\mathrm{Edd}$, the system is said to be in the intermediate state
(VHS/IS), where the central hot flow is still small in size and the flow remains
radiatively quite efficient. This state is also associated with a strong jet
\citep{Fender2004,Narayan2008,Fender2009}.

As the accretion rate decreases, the inner edge of the thin disk recedes further
outwards from the black hole (i.e. $r_\mathrm{tr}$ increases), resulting in a
larger inner flow which is very hot ($T_\mathrm{e} \sim 10^{9}-10^{11}$~K). For
$\dot{M} \sim 10^{-2}\dot{M}_\mathrm{Edd}$, the system is in the LHS, named for
its low luminosity and hard X-ray spectrum. The jet persists in this state,
albeit at a lower luminosity compared to the VHS/IS. Below $\dot{M} \sim
10^{-3}\dot{M}_\mathrm{Edd}$, the system enters the quiescent state --- a very
low luminosity state with a weak jet, where the system spends the majority of
its time \citep{Belloni2010}.

One common assumption that is used in the study of accretion flows is that there
should be equipartition between the magnetic and kinetic energy densities. This
assumption is made on the basis of scaling and energetics arguments, whereby the
magnetic energy density grows faster than the kinetic energy, reaching
equipartition quickly \citep[see, for example,][]{Shvartsman1971,Meszaros1975}.
Maintaining this equipartition then relies on the existence of dissipation
mechanisms in the disk, which subsequently act to keep the magnetic field at its
equipartition value.

The validity of this assumption has not been tested in-depth, although it is
generally relied on in many areas of accretion disk study \citep[see, for
example,][among others]{Melia1992,Narayan1996,Narayan1997a,Quataert2000,
Yuan2003,Bower2005,Beskin2005,Marrone2007,Kuo2014}. An initial attempt at
abandoning equipartition by \citet{Scharlemann1983} found that doing so would
introduce an instability to the flow, which could lead to an increased
dissipation rate of the magnetic field. The study was carried out for purely
spherical accretion flows and relied on a number of simplifying assumptions and
the theory was not developed for more general applications.
\citet{Kowalenko1999} discuss the effect of relaxing the equipartition
requirement, with the aim of incorporating turbulence in the flow into models of
the dissipation of magnetic field in a plasma. The authors find that at large
radii, the magnetic field may be suppressed to below its equipartition value,
while at small radii, the field may exceed equipartition, outstripping the
dissipation mechanism.

\added{Another study carried out by \citet{Karpov2001} found that under common
conditions for the formation of massive BHB systems, equipartition will not be
reached in a spherical accretion flow and as such, the magnetic field will
remain weak. This was followed by a recent paper \citep{Lipunov2021}, which
further discussed the evolution of BHB systems, again finding that a
quasi-spherical type of advection-dominated accretion is realized, with only a
weak magnetic field (below equipartition).} Despite the findings that removing
an equipartition prescription may have important implications for the magnetic
properties of the flow and for the heating of particles in the disk, this avenue
of investigation has not been pursued further, to our knowledge.

Numerical simulations by \citet{Igumenshchev2002,Igumenshchev2003} find that the
magnetic field may, in fact, be maintained at slightly sub-equipartition and
that it is dependent on the nature of the field supplied by the inflowing
matter. They also find that magnetic dissipation has a significant effect on the
structure of the flow, given that strong magnetic dissipation would contribute
to electron heating, as well as causing turbulence in the flow. It is clear then
that the assumption of equipartition may have significant effects on how the
flow evolves, in terms of heating and magnetic flux build-up.

Of the four main observational states in BHBs, three of the states (VHS/IS, LHS
and QS) are associated with the formation or presence of jets
\citep{Narayan2008}. In the intermediate and low/hard states, these jets are
powerful and dominate the emission in the radio/optical bands. In the quiescent
state, the jet power is substantially weaker and may no longer outshine the disk
emission. A central part of the study of relativistic jets is to understand the
role played by the magnetic field in launching the jets seen in these states
\citep{Blandford1977,Blandford1979,Casella2009,Tchekhovskoy2011}. It is
therefore vital to have a clear understanding of how the magnetic field behaves
in the inner regions of an accretion disk. In order to study the magnetic fields
within the flow, we turn to the quiescent state, where any contribution of a jet
to the spectrum is most suppressed and thus allows for closer examination of the
properties of the flow itself.

Previous studies have considered mainly the higher energy portions of the
spectrum, at and above $10^{15}$~Hz in particular. This was generally due to a
lack of observational data at lower frequencies (around $10^{10} - 10^{14}$~Hz),
where we expect to find synchrotron emission from the disk. Early attempts at
modelling BHB sources using a two-component accretion disk suggest that the
X-ray spectrum is formed by Comptonized disk photons and is found to be
insensitive to the exact magnetic properties of the disk
\citep[e.g.][]{Narayan1996,Narayan1997a,Esin1997,McClintock2003}. As a result,
it was not possible to constrain the magnetic field of the disk using these
observations.

More recent observations from ALMA, VLA and Spitzer have given more complete
coverage of the spectrum of some BHB sources, in particular at frequencies below
the peak of the thin disk emission, which occurs at around $10^{15}$~Hz
\citep{Narayan2008}. Sources such as V404 Cyg, XTE J1118+480 and A0620-00 have
been observed extensively at these frequencies
\citep{Gallo2007,Froning2011,Bernardini2016,Dincer2018,Gallo2019}, which makes
them ideal candidates to test for disk emission.

The exact source of the emission across the broadband spectrum is still the
subject of some debate. Most current models favour a jet-dominated or disk-jet
emission scenario to explain the broadband spectrum of the binary system, since
this can explain an observed correlation between radio and X-Ray luminosities
\citep{Gallo2003,Yuan2005}, while also providing an explanation for the flat or
inverted radio spectrum generally observed in quiescent sources
\citep{Hynes2009,Maitra2009,Froning2011,Xie2014,Russell2016,Yang2016}. On the
other hand, \citet{Veledina2013} argue that the optical and infrared
\added{(OIR)} parts of the spectrum can be explained by emission from an
extended hot accretion flow, rather than by emission from a jet. In particular,
the authors argue that the hot flow model should be favoured at these
frequencies since it can naturally explain the variability properties observed
in BHB sources.

In this paper, we propose a disk-dominated, two-component model, consisting of a
hot, roughly spherical flow in the inner regions of the system and a thin disk
in the outer regions, beyond the truncation radius $r_\mathrm{tr}$ in order to
probe the magnetic properties of the accretion disk. We want to find an
observational signature of the magnetic field within the flow and to this end,
we aim to measure the contribution of thermal synchrotron emission from the
inner regions of the quiescent disk to the overall spectrum of the source. We
apply this model to three sources, A0620-00, V404 Cyg and XTE J1118+480 in their
quiescent states, in order to examine the effect of changing the fraction of
equipartition, $\varepsilon_B$, which represents the ratio of the magnetic
energy density to the kinetic inflow energy density. We study the disk component
of the emission in the microwave, infrared and optical portions of the spectrum,
which show evidence of some excess emission \citep{Gallo2007}, commonly
attributed to a jet or outflow. In particular, we show for the first time that
the magnetic field present in the accretion flow is significantly weaker than
previously assumed, which conflicts with current theories of jet-launching and
suggests very strong magnetic dissipation in the flow.

\section{Model} \label{sec:model}

\begin{figure}[!htbp]
    \plotone{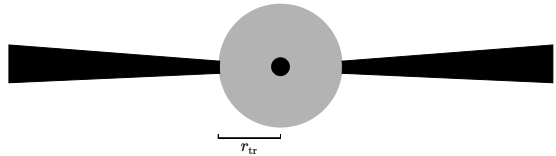}
    \caption{Schematic of the black hole and its surrounding accretion disk,
    showing the structure of the model used in this paper. The spherical gray
    area is the hot, inner flow, while the long, black regions on the outside
    represent the thin disk. The truncation radius, $r_\mathrm{tr}$ is the point
    at which the flow switches from thick to thin. Typical values of
    $r_\mathrm{tr}$ are $10^3 - 10^4 r_g$. The overall system size is $\sim 10^5
    r_g$.}
    \label{fig:model}
\end{figure}

Our model consists of two components, namely an inner, hot flow and and an
outer, thin disk (a schematic of this model is shown in figure~\ref{fig:model}).
The inner flow radiates via thermal synchrotron and bremsstrahlung emission,
while the thin disk is modelled as an MCD. This model is supported by
observational evidence, which suggests that the thin disk does indeed retreat
from the inner regions in the lower luminosity states, particularly in the
quiescent state, where there is no evidence for the soft blackbody emission at
X-Ray wavelengths that would be expected from the inner regions of a thin disk
\citep{Zdziarski2004,Done2007}.

\subsection{Thin Disk} \label{subsec:thindisk}

The spectrum of the outer disk is constrained by the mass of the central black
hole in the system and the accretion rate. For a given frequency $\nu$, the
specific emitted flux, $F_\nu$, is \citep{Frank2002}:
\begin{equation}
F_\nu = \frac{4 \pi h \cos(i) \nu^3}{c^2 d^2}
\int^{r_\mathrm{out}}_{r_\mathrm{in}}\frac{r\mathrm{d}r}{\exp(h\nu/k_BT(r))-1},
\label{eq:thinflux}
\end{equation}
where:
\begin{equation}
T(r) = \left\lbrace \frac{3GM\dot{M}}{8\pi r^3 \sigma}
\left[1 - \left(\frac{r}{r_s}\right)^{1/2}\right] \right\rbrace^{1/4}.
\label{eq:thintemp}
\end{equation}

Here, $i$ is the inclination of the thin disk to the line of sight, $M$ is the
mass of the central object, $\dot{M}$ is the accretion rate, $r_s$ is the
Schwarzschild radius and $d$ is the distance from the source to the observer.
$G$ is the gravitational constant, $h$ is Planck's constant, $\sigma$ is the
Stefan-Boltzmann constant and $c$ is the speed of light.

The value of $M$ can in general be determined independently of the spectrum
(e.g. by dynamical measurement), so for sources where this is the case, it is a
fixed parameter of this model.

For sources with independent mass measurements, the accretion rate, $\dot{M}$,
is generally constrained by the contribution of the thin disk to the spectrum
--- since the temperature of the disk is dependent only on mass and accretion
rate. Matching the observed flux can therefore set the constraint on the mass
inflow rate, which is also important for the emission from the thick disk, since
the thin disk feeds directly into the inner flow. The other parameters that
affect the spectrum of the thin disk are the values of $r_\mathrm{in}$,
$r_\mathrm{out}$ and the inclination of the thin disk to the line of sight, $i$.

$r_\mathrm{in}$ in this case corresponds to the truncation radius,
$r_\mathrm{tr}$. Typical values of $r_\mathrm{tr}$ are expected to be a few
$10^3 r_g$ to $\sim 10^4 r_g$ \citep{Narayan2008}. Values of $r_\mathrm{out}$
for these systems generally lie at around $10^5 r_g$, due to the fact that for
an accreting binary system, the in-falling material must come from the companion
star and so the size of its accretion disk cannot be larger than the
\replaced{orbital size}{Roche lobe} of the system.

The value of the inclination is in many cases less certain, although there are
some observational constraints for a limited number of sources.

\subsection{Thick Disk} \label{subsec:thickdisk}

The contribution of the inner disk is mainly in the form of synchrotron
emission, where a thermal population of electrons is assumed. We tie the
magnetic field in the flow to the kinetic energy of the infalling material by
assuming that the magnetic energy density is in some fraction of equipartition,
$\varepsilon_B$ with the kinetic infall energy, \added{defining $\varepsilon_B =
u_m / u_k$, where $u_m$ is the magnetic energy density and $u_k$ is the kinetic
energy density} \citep[following][for
example]{Shvartsman1971,Melia1992,Marrone2006}.

The resulting model of the inner accretion flow is as follows: The radial
density profile is assumed to follow a power law:
\begin{equation}
n(r) = n_0\left(\frac{r}{r_s}\right)^{-\beta},
\label{eq:density}
\end{equation}
where here $n_0$ is the number density at the Schwarzschild radius. The value of
$\beta$ is determined by the structure of the accretion flow. For free-falling
gas \citep{Blandford1999}, $\dot{M}(r) \propto r^p$, and then $\beta =
\frac{3}{2}-p$. For spherical accretion \citep{Bondi1952} or an advection
dominated accretion flow (ADAF) \citep{Narayan1994}, $\beta = \frac{3}{2}$,
while for a convection-dominated accretion flow (CDAF), $\beta = \frac{1}{2}$.
In this work, the value of $\beta$ is taken to be 1.5, corresponding to an
ADAF/Spherical flow.

\replaced{Some fraction of equipartition is assumed between magnetic, kinetic
and gravitational energy}{The equipartition ratio $\varepsilon_B$ is used to
vary the strength of the magnetic field in our model, where a value of
$\varepsilon_B = 1$ corresponds to equipartition between the kinetic and
magnetic energy densities, while values of $\varepsilon_B < 1$ represent a
sub-equipartition magnetic field (i.e. lower magnetic field strength)}
\citep[e.g.][]{Shvartsman1971, Melia1992}. This allows the number density of
particles $n_e$ to be tied to the magnetic field strength $B$. Then making use
of the equation for the density as a power law and assuming pure hydrogen in the
plasma gives: $\rho = m_\mathrm{H} n_0\left(\frac{r}{r_s}\right)^{-\beta}$.
Equating the energy densities results in: $\frac{B^2}{8 \pi} =
\varepsilon_B\frac{1}{2}\rho u_r^2,$ where $u_r$ is the infall speed, with $u_r
\sim \frac{c}{\sqrt{\frac{r}{r_s}}}$ \citep[e.g.][]{Narayan2008}. Thus the
magnetic field strength is:
\begin{equation}
    B = \sqrt{4\varepsilon_B \pi c^2 m_{\mathrm{H}}n_0}
    \left(\frac{r}{r_s}\right)^{\frac{-(\beta +1)}{2}}.
\label{eq:magnetic_fs}
\end{equation}

The temperature at the inner edge of the accretion flow is a free parameter of
the model, with typical electron temperatures of $\sim 10^9 - 10^{11}$K for a
two-temperature flow \citep{Narayan2008}.

We use the angle-averaged synchrotron emissivity from \citet{Wardzinski2000}:
\begin{equation}
j_\nu =
\frac{2^{1/6}\pi^{3/2} e^2 n_e \nu}{3^{5/6} c K_2\left(1/\Theta\right) v^{1/6}}
\exp\left[-\left(\frac{9v}{2}\right)^{1/3}\right],
\label{eq:emissivity}
\end{equation}
with $\Theta = kT/mc^2$ and $v = \nu/\left(\nu_c\Theta^2\right)$, $\nu_c$ is the
cyclotron frequency. Here, $K_2$ is the modified Bessel function of the second
kind \citep[e.g.][]{Abramowitz1970}.

For thermal electrons, the absorption coefficient is given by Kirchhoff's Law:
$\alpha_\nu = j_\nu / B_\nu$ \citep{Rybicki1979}. To find the specific
intensity, we solve numerically the radiative transfer equation along the line
of sight:
\begin{equation}
\frac{\mathrm{d}I_\nu(s)}{\mathrm{d}s} = -\alpha_\nu(s) I_\nu(s) + j_\nu(s).
\label{eq:radiative-transfer}
\end{equation}
To calculate the specific flux from a spherical source,
\citep[see][figure~1.6]{Rybicki1979}, we get:
\begin{equation}
F_\nu = \pi \int_{s_0}^{s}{I_\nu(s')\sin\theta_c(s')^2 \mathrm{d}s'},
\label{eq:specific-flux-intermediate}
\end{equation}
where: $\theta_c(s) = \sin^{-1}(s/d)$ (for an observer at distance $d$). Thus
the final expression for the specific flux is:
\begin{equation}
F_\nu = \frac{\pi}{d^2}\int_{s_0}^{s}{I_\nu(s')s'^2 \mathrm{d}s'}.
\label{eq:specific-flux-solution}
\end{equation}

\section{Modelled Sources} \label{sec:sources}

\subsection{A0620-00} \label{subsec:a0620}

A0620-00 is a soft X-Ray transient source, consisting of a $6.6~M_\odot$ compact
object and a $0.4~M_\odot$ K-type companion, with an orbital period of 7.75~hr
\citep{McClintock1986,Cantrell2010}. It is one of the most extensively observed
BHB systems to date. The inclination of the system is estimated at: $i =
\left(51 \pm 0.9\right)^\circ$, while the distance from Earth is $d = \left(1.06
\pm 0.12\right)$~kpc \citep{Tetarenko2016}.

In constructing the broadband \replaced{SED}{spectral energy distribution (SED)}
of A0620-00, we use the above parameters for the central black hole, along with
an accretion rate of $2.75\times10^{-12}~M_\odot$ per year. This rate is
initially based on previous estimates, e.g. \citet{Narayan1996}, and chosen to
match the thin disk contribution to the spectrum.

Furthermore, we adopt an outer disk radius, $r_\mathrm{out}$, of $10^5~r_g$, on
the basis that the orbital period implies a system size of around $10^{11}$~cm,
which corresponds to $\sim 10^5~r_g$. The value of $r_\mathrm{tr}$ is varied
along a few $10^3 - 10^4~r_g$, which is in line with standard choices by other
modellers. The value of $r_\mathrm{tr}$ is less well constrained and is
generally treated as a free parameter of most models. \citet{Narayan1996} offer
a possible method for choosing a reasonable value, based on comparisons with the
H$\alpha$ line, although the authors find that an accurate determination using
this method is tricky and maintain $r_\mathrm{tr}$ as a free parameter.

We compare with observations as reported in \citet[][and references
therein]{Gallo2019}, which give broadband coverage of the spectrum from radio to
X-Ray frequencies. In particular, we focus on explaining the excess of emission
observed at OIR frequencies, where the shape of the spectrum deviates from that
expected of a standard MCD blackbody. We find that this component of the
emission can be explained as thermal synchrotron emission coming from the inner
flow.

In the case of A0620-00, the addition of ALMA data allows us to further
constrain the properties of the disk, which then enables us to estimate the
value of $\varepsilon_B$ in the flow.

\subsection{XTE J1118+480} \label{subsec:xtej1118}

XTE J1118+480 is another soft X-Ray transient source, consisting of a
$7.3~M_\odot$ compact object and a $0.2~M_\odot$ companion, with an orbital
period of 4.1~hr. The inclination of the system is in the range:
$68^\circ~<~i~<82^\circ$, while the distance from Earth is $d = \left(1.72 \pm
0.1\right)$~kpc \citep{Tetarenko2016}.

We compare with observations as reported in \citet[][and references
therein]{Gallo2007}. Again focusing on explaining the excess of emission
observed at OIR frequencies, which is exhibited by XTE J1118+480 as in A0620-00.

Here, the orbital period again implies a system size of around $\sim
10^{11}$~cm, which corresponds to $\sim 10^5~r_g$. The value of $r_\mathrm{tr}$
is chosen to be $5~\times~10^3~r_g$, along with an accretion rate of
$3\times10^{-12}~M_\odot$~yr$^{-1}$.

\subsection{V404 Cygni} \label{subsec:v404}

V404 Cyg consists of a \replaced{$7.1~M_\odot$}{$9~M_\odot$}\explain{This and
further corrections to the parameters of V404 Cygni are in response to feedback
from ArXiv posting.} compact object and a $0.7~M_\odot$ K-type companion, with
an orbital period of 155~hr \citep{Bernardini2016}. The inclination of the
system is estimated at: \replaced{$i = \left(81 \pm 0.9\right)^\circ$}{$i =
\left(67^{+3}_{-1}\right)^\circ$ \citep{Khargharia2010}}, while the distance
from Earth is: \replaced{$d = \left(1.06 \pm 0.12\right)$~kpc}{$d = \left(2.39
\pm 0.14\right)$~kpc} \citep{Tetarenko2016}. The accretion rate in V404 Cyg is
less certain, however it can be estimated using the shape of the blackbody
spectrum (as discussed in section~\ref{subsec:thindisk}). Using this procedure
to set the accretion rate leads to a value of
\replaced{$10^{-9}~M_\odot$}{$1.5\times10^{-9}~M_\odot$} per year.

We adopt an outer disk radius, $r_\mathrm{out}$, of $10^5~r_g$, as the orbital
period again implies a system size of around $10^{12}$~cm, which corresponds to
$\sim 10^5~r_g$ for a \replaced{$7~M_\odot$}{$9~M_\odot$} black hole. The value
of $r_\mathrm{tr}$ is varied as before, along a few $10^3 - 10^4~r_g$.

V404 Cyg also displays the excess of emission at OIR frequencies below the thin
disk peak, which we argue can again be explained as thermal synchrotron emission
from the inner flow. The lack of spectral coverage at wavelengths between the
radio and the OIR means that it is not possible to confidently constrain the
magnetic field in V404 Cyg's inner accretion flow, although models with
sub-equipartition fields can produce the necessary excess emission and so
further observations (e.g. with ALMA) could place stronger constraints on the
field.

\section{Results} \label{sec:results}

It is important to note that in each of these cases, we are not attempting a
statistical best fit, but rather to generate a representative spectrum which
visually reproduces the main features of each spectrum and illustrates the
magnitudes of the quantities involved. These results should therefore not be
taken as an attempt to determine the exact parameters of the sources in
question, but rather to demonstrate the properties of the magnetic field in our
model in a theoretical sense, making use of values already available in the
literature, such as accretion rate and inclination.

\begin{figure}[!htbp]
    \plotone{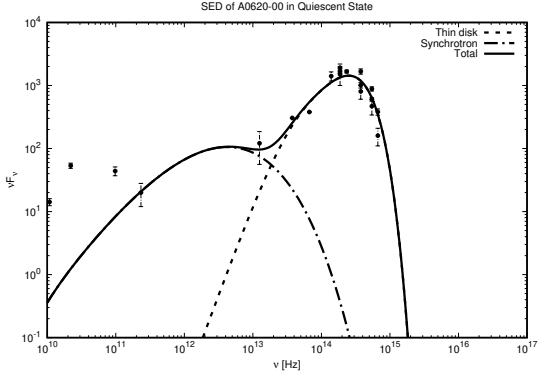}
    \caption{SED of A0620-00, along with observational data from \citet[and
    references therein]{Gallo2019}. The excess of emission near $10^{13}$~Hz is
    well described by thermal synchrotron from a weakly-magnetized, hot flow.
    The radio data are not explained by the synchrotron portion of the spectrum,
    suggesting that some contribution from a radio emitting jet or outflow would
    still be necessary to fully explain the observed spectrum of A0620-00.}
    \label{fig:a0620full}
\end{figure}

\begin{figure}[!htbp]
    \plotone{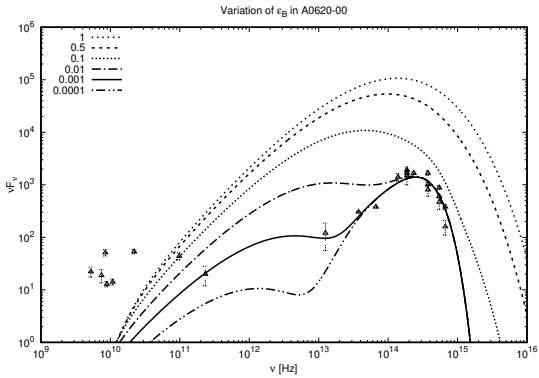}
    \caption{The effect of changing the value of $\varepsilon_B$, showing that
    an equipartition magnetic field overproduces emission and dominates the
    contribution from the thin disk. Moving to lower values of $\varepsilon_B$
    results in the characteristic ``bump'' in the spectrum at around
    $10^{13}$~Hz. Clearly, the value of the magnetic field must be significantly
    below equipartition in order to explain the observed excess emission.}
    \label{fig:varyeq}
\end{figure}

In the case of A0620-00, we find good qualitative agreement between our model
and the data for a very weak magnetic field (see figure~\ref{fig:a0620full}).
For the parameters given in section~\ref{subsec:a0620}, we find that an
accretion disk with a temperature of $6\times10^{10}$~K at the inner edge of the
accretion disk and a truncation radius of $4\times10^{3}~r_g$ can well explain
the observed excess radiation with an equipartition ratio $\varepsilon_B =
10^{-3}$.

Figure~\ref{fig:varyeq} clearly illustrates that the excess emission at OIR
frequencies is naturally explained by a sub-equipartition model. We disfavour
models with values of $\varepsilon_B$ close to equipartition, since these models
overproduce emission at all frequencies between $10^{11}$~Hz and $10^{15}$~Hz.

The two-component model cannot explain the emission at radio frequencies for any
sensible choice of parameters, with a sharp cut-off in the synchrotron emission
component occurring at around $10^{10}$~Hz, meaning that some contribution from
a jet or outflow would have to be present to account for the lower energy
radiation.

\begin{figure}[!htbp]
    \plotone{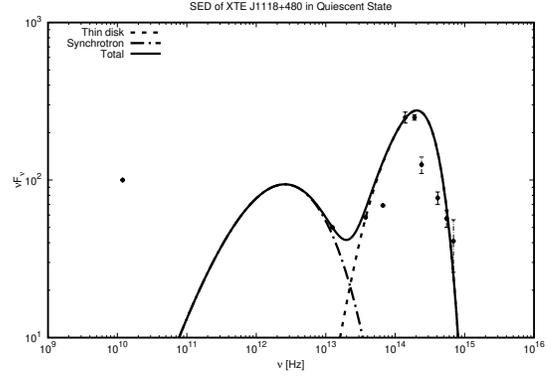}
    \caption{SED of XTE J1118+480, along with observational data from \citet[and
    references therein]{Gallo2007}. The excess of emission near $10^{13}$~Hz is
    once again well described by thermal synchrotron from a weakly-magnetized,
    hot flow.}
    \label{fig:xtej1118full}
\end{figure}

The excess OIR emission in XTE J1118+480 can similarly be modelled as thermal
synchrotron emission coming from a weakly magnetized inner flow (see
figure~\ref{fig:xtej1118full}). Using the values in
section~\ref{subsec:xtej1118}), combined with an inner temperature of
$5\times10^{10}$~K and a truncation radius of $5\times10^{3}~r_g$, there is good
agreement with the observed data points if we again choose $\varepsilon_B =
8\times10^{-4}$ --- a very weak field.

\begin{figure}[!htbp]
    \plotone{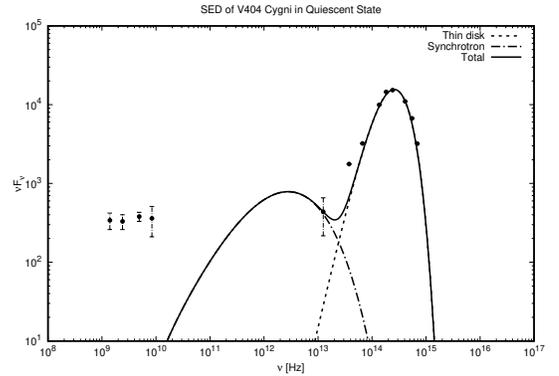}
    \caption{SED of V404 Cyg, along with observational data from \citet[and
    references therein]{Gallo2007}. The excess of emission near $10^{13}$~Hz is
    well described by thermal synchrotron from a weakly-magnetized, hot flow.
    The parameters used in this case are: $T = 3\times10^{11}$~K, $r_\mathrm{tr}
    = 3\times10^{4}~r_g$. The value of $\varepsilon_B$ is found to be
    \replaced{$10^{-9}$}{$2\times10^{-9}$}, which is extremely small, even when
    compared with that of the previous two sources.}
    \label{fig:v404full}
\end{figure}

In order not to overproduce synchrotron emission in V404 Cyg, we require a value
of \replaced{$\varepsilon_B = 10^{-9}$}{$\varepsilon_B = 2\times10^{-9}$},
corresponding to a magnetic field strength of around 50~G at $10~r_g$, which is
far below equipartition and far below the value of $\varepsilon_B$ found in the
previous two sources (see~\ref{fig:v404full}). Indeed, in the case of A0620-00,
our model suggests a magnetic field strength of around 2000~G at a distance of
10~$r_g$ from the central black hole. For XTE J1118+480, we calculate a value of
1750~G. We therefore find that the magnetic field in V404 Cyg must be
significantly smaller than that in the other two sources, according to this
analysis.

One possible explanation for this difference between sources is that the
accretion rate in V404 Cyg is much higher than in the other two sources
($\dot{M}~\sim~10^{-9}$ vs. $\dot{M}~\sim~10^{-12}$). This higher accretion rate
should still correspond to a system in the quiescent state, given that it
represents an accretion rate of around $10^{-2}\dot{M}_\mathrm{Edd} -
10^{-3}\dot{M}_\mathrm{Edd}$ \citep[see the analysis by][for
example]{Narayan2008}. It should be noted however, that this value does lie on
the upper limit of the range of accretion rates expected in the quiescent state
and so the case of V404 Cyg may warrant further investigation in order to
determine exactly which accretion state it is exhibiting.

\section{Discussion} \label{sec:discussion}

The observational data for all these sources is combined from several different
observing campaigns, sometimes spanning several years. While all of the data
points we use correspond to observations of the sources in their quiescent
states, the nature of BHBs is that they are variable in a number of parameters,
most notably accretion rate. This can have significant observational
consequences, since the accretion rate is closely tied to the behaviour of key
aspects of the system. In particular, the flux coming from the thin disk is
essentially entirely dependent on the value of $\dot{M}$ (where the mass of the
central object is known). This combines with the understanding that any
variation in the accretion rate will result in a change in the truncation
radius. The magnitude of this change is not clear, however it is important to
bear in mind when assessing these results. Observations of A0620-00 show that
the accretion rate appears to remain somewhat stable over time, which lends
weight to the assumption that the overall shape of the spectrum will remain
largely unchanged over time (as long as it is observed in the same spectral
state).

Sub-equipartition models provide a good explanation for the emission from the
all three sources, although only in the case of A0620-00 do we have enough data
to rule out some of the higher temperature cases. With this constraint, the
value of the magnetic field falls to as low as 0.1\% of the equipartition value.
In all cases, for reasonable choices of parameters, an equipartition magnetic
field would overproduce emission at sub-millimeter wavelengths. As mentioned
previously, the lack of simultaneous data in all of the regions of interest may
present an obstacle to making firm conclusions, however since the accretion rate
can be constrained by the thin disk component and since all of the sources show
an excess in the OIR region, it is reasonable to assume that it should exist as
a general feature of the spectrum across long time periods.

The value of the magnetic field in V404 Cyg was found to be extremely small,
even relative to the small fields found in A0620-00 and XTE J1118+480. Some part
of this discrepancy may be explained by the higher accretion rate in V404 Cyg,
which lies towards the upper limit of accretion rates believed to be found in
the quiescent state, at around 0.1\%--1\% of the Eddington value. This situation
may not be well described by the simple two-component model used in this work.
In reality, the dynamics of the disk are likely to be significantly more complex
and it stands to reason that the largest deviations from this simplified
behaviour would occur for extreme values of the quiescent accretion rate.

If the disk geometry is significantly altered by a recent transition from a
higher luminosity state (e.g. due to the presence of a strong jet or outflow,
which may disrupt or otherwise alter the inner regions of the accretion flow),
numerical simulations of the system may be required to constrain the accretion
mode and to assess whether or not the system has fully settled into the
quiescent state. There are some hints that V404 Cyg's accretion mode and
geometry may not be identical to that of the other two sources, given that the
accretion rate of V404 Cyg is higher, however we require a significantly larger
truncation radius in order to fit the blackbody portion of the spectrum. This
runs counter to what is usually expected, whereby a higher accretion rate
corresponds to a smaller transition radius. This is possibly due to some effects
of state transitions in the system, which should coincide with the shutting off
of a powerful jet as the system moves into the quiescent state. Over time, it is
also expected that the outer disk will begin to move inwards, eventually leading
to the reappearance of the jet.

The large value of the truncation radius used in this case will result in the
temperature of the inner disk dropping quite low. This will have some
implications on the structure of the disk, however the ADAF solution will still
be valid in this region and so should not pose a major problem for this
analysis. The drop in temperature does present a larger problem however, which
is discussed in detail in \citet{Wardzinski2000}, namely that as the
dimensionless temperature $\Theta$ falls below a value of around $0.9$, the
emissivity (equation~\ref{eq:emissivity}) will tend to overestimate the rate of
production of synchrotron emission. This issue can be partially alleviated by
the physical argument that the disk emission should be largely dominated by
emission from the innermost regions, where the temperature, density and magnetic
field are highest. As such, ignoring the effects of equation~\ref{eq:emissivity}
beyond around $10^{4}~r_g$ should not be a drastic departure from the true
value. Direct numerical solutions of the synchrotron emission, based on
numerical disk simulations, will provide a more comprehensive analysis of
extreme cases such as this, without relying on analytical approximations of the
emissivity to the same degree.

\begin{figure}[!htbp]
    
    \plotone{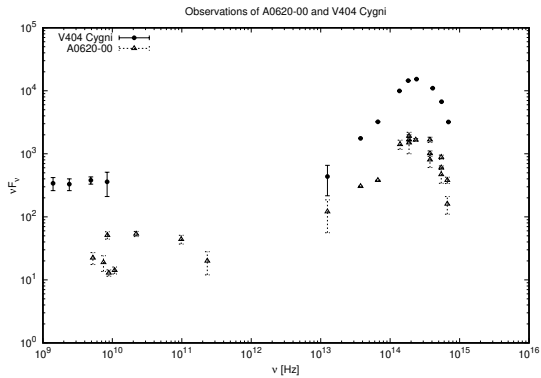}
    \caption{Comparison of the observed fluxes in V404 Cyg (circles) and
    A0620-00 (triangles), which shows that the difference in synchrotron flux is
    a factor of a few. The most relevant area of comparison based on our
    calculation is the point around $10^{13}~\mathrm{Hz}$\deleted{`'}, which
    comes from the synchrotron emission in the inner flow. The higher energy
    emission comes from the outer thin disk in our model, while the lower energy
    emission is likely to come from a jet or outflow.}
    \label{fig:fluxcomparison}
\end{figure}

The low value of $\varepsilon_B$ can, however, be verified by comparing the
total synchrotron power of V404 Cyg with one of the other sources. Given that:
\begin{equation}
    P_\mathrm{tot} \sim n B^2 T^2
    \label{eq:total-power}
\end{equation}
\citep[e.g.][]{Wardzinski2000}, comparing the relevant quantities in V404 Cyg
and A0620-00, we see that a value of \replaced{$\varepsilon_B =
10^{-9}$}{$\varepsilon_B = 2\times10^{-9}$} in V404 Cyg produces:
\begin{equation}
    \frac{P_\mathrm{V}}{P_\mathrm{A}} =
    \frac{n_\mathrm{V}}{n_\mathrm{A}}
    \frac{B^2_\mathrm{V}}{B^2_\mathrm{A}}
    \frac{T^2_\mathrm{V}}{t^2_\mathrm{A}}
    \sim 4.5.
    \label{eq:powercomparison}
\end{equation}
Plotting the observed fluxes together (see figure \ref{fig:fluxcomparison})
indicates that the expected power in our calculations is correct as expected by
standard synchrotron theory.

Indeed, we find agreement with the calculations of \citet{Dallilar2017}, who
find a magnetic field of 33~G from synchrotron fitting, which matches
\replaced{almost exactly}{very closely} with the 50~G we find. The authors also
agree that the source is likely to be in a non-equipartition configuration.
Similarly, \citet{Jana2020} find a magnetic field in V404 Cyg of 90--900~G,
although their analysis deals with the outburst phase of the source and uses a
slightly different model, in which the thin disk extends all the way to the ISCO
and is surrounded by the optically thin region, rather than replaced by it. In
any case, the important finding is that the magnetic field strength is likely to
be low in this source.

A sub-equipartition magnetic field suggests that any dissipation mechanism of
the magnetic field is likely to be quite strong, providing significant heating
to the accretion flow. Given that the magnetic field grows faster than the
kinetic energy, such a dissipation mechanism must exist and furthermore, must
operate efficiently. There are numerous potential instabilities that can exist
within an accretion disk, many of which will work to dissipate the field.
Previously, the assumption of equipartition relied on these mechanisms favouring
a magnetic field at its equipartition value and working constantly to maintain
such a scenario. If it is the case that magnetic fields do not tend towards
equipartition, then there are some consequences that should be considered.

Firstly, if the value of the $\varepsilon_B$ is not observed to be constant over
a large number of sources, it is possible that the dissipation of magnetic field
follows a random process. This is certainly possible, given the turbulent nature
of accretion flows, although it is not clear that it would be energetically
favourable.

If, on the other hand, $\varepsilon_B$ is observed to follow a clear relation
across different accreting sources, then the fundamental idea of an
equipartition assumption need not change. That is, if the magnetic field
generally tends to a particular value of $\varepsilon_B$, then the notion of a
dissipation mechanism within the flow acting to maintain a particular magnetic
configuration would still hold true, albeit for a significantly smaller magnetic
field strength than previously assumed. The energetics of such a situation may
suggest the presence of another balancing quantity, but such considerations are
beyond the scope of this paper.

If the accretion disk cannot, or rather does not sustain a large magnetic field,
this could pose a problem for the launching of jets from the inner accretion
flow, which is generally understood to require the presence of strong magnetic
fields. Given that the two-component model in this paper cannot adequately
explain the observed flat spectrum at low energies, which is characteristic of a
self-absorbed jet or outflow, it seems likely that some form of jet is present
to account for this emission. This jet would have to be quite weak, with a break
coming at relatively low frequencies, to avoid contaminating the disk
contribution. Avoiding sub-equipartition fields in the accretion flow possibly
requires substantial mass loss between the outer edge of the disk and the
innermost regions, resulting in a significantly lower accretion rate at the
horizon of the black hole, compared to that at $r_\mathrm{out}$.

It is unclear whether this configuration is unique to the quiescent state, since
it is not possible to directly probe disk emission in the LHS and there is no
thick disk component in the HSS. Discontinuities in the magnetic field structure
across states do not seem likely, given the more-or-less continuous behaviour of
other system parameters between states, however the influence of a jet on the
overall magnetic structure of the system (and vice-versa) is still not a settled
question and may provide some clues as to what happens between states.

\section{Conclusions} \label{sec:conclusions}

In this paper, we have modelled the SEDs of 3 BHB sources in order to examine
the common assumption of equipartition between the magnetic and kinetic energies
in the accretion flow. We use a two-component model, consisting of a hot,
optically thin, geometrically thick inner flow, surrounded by a cold, optically
thick, geometrically thin outer flow to reproduce the main spectral features of
the sources (we do not perform statistical fits for this purpose). We find that
an excess of emission observed at OIR frequencies can be explained as thermal
synchrotron emission coming from the hot inner region of the accretion flow. We
argue for a magnetic field that is sub-equipartition, with values of
$\varepsilon_B$ well below unity, running as low as $10^{-9}$ in the case of
V404 Cyg. This is much lower than often assumed and has implications for the
mechanisms of dissipation of magnetic fields within accretion flows, as well as
the launching of jets that are often associated with BHB sources. In particular,
many current jet models rely on a large magnetic field to power the jet. The
absence of such a large magnetic field could prove problematic to a number of
current theories surrounding jet-formation, which warrants further investigation
to reconcile these points.

This result underscores the importance of OIR observations of BHBs, which can
provide important constraints on the strength of the magnetic field strength in
the inner accretion flow. Application of this model to further sources in the
quiescent state may reveal whether the low magnetic field holds across a variety
of sources, or if there is significant variation in magnetic regimes between
sources. In either scenario, there are important implications for our
understanding of the role of the magnetic field in launching jets.

\added{In combination with other tests, such as those presented by
\citet{Cherepashchuk2019b,Cherepashchuk2019a}, which use photometric testing to
model light curves and construct spectra of binary systems in their quiescent
state, these results could provide a useful test for theoretical models of
accretion in binary systems. Any successful model of a binary system would need
to adequately explain both the spectral features observed, as well as the
temporal features of the light curve. This could allow for the exclusion of
models that are either spectrally or temporally inconsistent with observational
data, providing an important observational diagnostic for modellers.}

Numerical studies of weakly-magnetized disks should provide insight into whether
such configurations are consistent with jet-launching. In addition, numerical
studies may provide a differentiator between the LHS and QS, which are sometimes
understood to be close relatives, due to their structure (large, thick, inner
disk, surrounded by outer thin disk). If the magnetic configuration in the QS
proves to be different from that in the LHS, it may account for the difference
in jet power observed in each state. If the LHS is well explained with a low
magnetic field configuration as in the QS, it may be that the role of the
magnetic field in powering the jet is less important than previously believed.

\begin{acknowledgments}
This paper makes use of the following ALMA data: ADS/JAO.ALMA\#2016.1.00773.S.
ALMA is a partnership of ESO (representing its member states), NSF (USA) and
NINS (Japan), together with NRC (Canada), MOST and ASIAA (Taiwan), and KASI
(Republic of Korea), in cooperation with the Republic of Chile. The Joint ALMA
Observatory is operated by ESO, AUI/NRAO and NAOJ. The National Radio Astronomy
Observatory is a facility of the National Science Foundation operated under
cooperative agreement by Associated Universities, Inc. This research has made
use of NASA's Astrophysics Data System Bibliographic Services.

AP acknowledges support from the European Research Council via the ERC
consolidating grant \#773062 (acronym O.M.J.).
\end{acknowledgments}

\vspace{5mm}
\facilities{ALMA, CTIO:1.3m (ANDICAM), CXO, Spitzer, VLA}



\bibliography{equipartition}
\bibliographystyle{aasjournal}

\end{document}